\documentclass{article}
\usepackage{amsfonts}
\usepackage{amsmath}

= -2truecm

\begin{document}

\title{Generating superposition and entanglement of squeezed vacuum states}
\author{Zhi-Ming Zhang\thanks{%
zmzhang@scnu.edu.cn} \\
School of Information and photoelectronic Science \& Technology\\
South China Normal University, Guangzhou 510006, China}
\maketitle

\begin{abstract}
We propose a scheme for generating the superposition and the entanglement of
squeezed vacuum states of electromagnetical fields. The scheme involves
single-photon source, single photon detector, and cross Kerr nonlinearity.
The Kerr nonlinearity required for generating the superposition of squeezed
vacuum states is $1/2$ of that required for generating the superposition of
coherent states. The proposal can also be extended to generate the
entanglement states between two coherent states and that between one
coherent state and one squeezed vacuum state.
\end{abstract}

Superposition and entanglement are characteristic features of quantum states
and have important applications in quantum science and technology, for
example, in quantum computation and quantum information \cite{Nielsen}. The
superpositions and entanglements of coherent states have been studied
extensively\cite{Cat}. Especially, there have been experimental
demonstrations of generating the superpositions of coherent states in a
cavity \cite{Haroche} and in an ion trap \cite{Wineland}. However, few
attention has been paid on the study of superposition and entanglement of
squeezed states\cite{Kitagawa}. Recently, Chen et al.\cite{Chen} proposed a
scheme for generating superposition and entanglement of squeezed states, and
their scheme was based on cavity quantum electrodynamics (cavity QED). In
this paper, we propose a simple scheme for generating superposition and
entanglement of squeezed vacuum states. Our work was inspired by Gerry's idea%
\cite{Gerry}, and involves single photon source, single photon detector,
beam splitter, phase-shifter, and cross Kerr nonlinearity. The scheme is
shown in Fig.1. A photon in mode $b$ is incident on the beam splitter BS1.
One of the output modes of\ BS1 interacts with mode $a$ (in a squeezed
state) via a cross Kerr nonlinearity in the Kerr medium KM, and the other
output mode of\ BS1 passes through the phase-shifter $\theta $, then the two
modes combine at the beam splitter BS2. By detecting output photons from the
beam splitter BS2, the states of mode $a$ will be projected into a
superposition of squeezed states. By changing the configuration of Fig.1 to
that of Fig.2, we can obtain entanglement of squeezed states. Now we discuss
the schemes in more details.

We assume the beam splitters are of the type of 50:50, as shown in Fig.3,
then we have%
\begin{equation}
\begin{array}{c}
b_{out}=\frac{1}{\sqrt{2}}\left( b_{in}+ic_{in}\right) , \\
c_{out}=\frac{1}{\sqrt{2}}\left( c_{in}+ib_{in}\right) ,%
\end{array}
\label{1}
\end{equation}%
or
\begin{equation}
\begin{array}{c}
b_{in}^{+}=\frac{1}{\sqrt{2}}\left( b_{out}^{+}+ic_{out}^{+}\right) , \\
c_{in}^{+}=\frac{1}{\sqrt{2}}\left( c_{out}^{+}+ib_{out}^{+}\right) ,%
\end{array}
\label{2}
\end{equation}%
where $b\left( b^{+}\right) $ and $c\left( c^{+}\right) $ are the photon
annihilation (creation) operators of mode $b$ and mode $c$, respectively.
The beam splitter has following properties%
\begin{equation}
\begin{array}{c}
\left\vert 1\right\rangle _{b}\left\vert 0\right\rangle _{c}\rightarrow
\frac{1}{\sqrt{2}}\left( \left\vert 1\right\rangle _{b}\left\vert
0\right\rangle _{c}+i\left\vert 0\right\rangle _{b}\left\vert 1\right\rangle
_{c}\right) , \\
\left\vert 0\right\rangle _{b}\left\vert 1\right\rangle _{c}\rightarrow
\frac{1}{\sqrt{2}}\left( \left\vert 0\right\rangle _{b}\left\vert
1\right\rangle _{c}+i\left\vert 1\right\rangle _{b}\left\vert 0\right\rangle
_{c}\right) .%
\end{array}
\label{3}
\end{equation}%
The phase-shifter operator has the form $U_{PS}\left( \theta \right) =e^{i%
\hat{n}\theta },$and the property%
\begin{equation}
U_{PS}\left( \theta \right) \left\vert n\right\rangle =e^{in\theta
}\left\vert n\right\rangle ,  \label{4}
\end{equation}%
in which $\hat{n}$ is the photon-number operator, and $\left\vert
n\right\rangle $ is its eigenstate. The interaction Hamiltonian for a cross
Kerr nonlinearity has the form%
\begin{equation}
H_{CK}=\hbar K\hat{n}_{a}\hat{n}_{b},  \label{5}
\end{equation}%
where $\hat{n}_{a}=a^{+}a$ and $\hat{n}_{b}=b^{+}b$ are the photon-number
operator of mode $a$ and mode $b$, respectively. $K$ is proportional to the
third-order nonlinear susceptibility $\chi ^{\left( 3\right) }$. The
time-evolution operator is

\begin{equation}
U\left( t\right) =\exp \left( -\frac{i}{\hbar }H_{CK}t\right) =\exp \left\{
-iK\hat{n}_{a}\hat{n}_{b}t\right\} =\exp \left\{ -i\tau \hat{n}_{a}\hat{n}%
_{b}\right\} =U\left( \tau \right) ,  \label{6}
\end{equation}%
in which $\tau =Kt=K\left( l/v\right) $, here $l$ is the length of the Kerr
medium and $v$ is the velocity of light in the Kerr medium. The
time-evolution operator has the following properties
\begin{equation}
U\left( \tau \right) \left\vert n\right\rangle _{b}\left\vert \alpha
\right\rangle _{a}=\left\vert n\right\rangle _{b}\left\vert \alpha
e^{-in\tau }\right\rangle _{a},  \label{7}
\end{equation}%
\begin{equation}
U\left( \tau \right) \left\vert n\right\rangle _{b}\left\vert \xi
\right\rangle _{a}=\left\vert n\right\rangle _{b}\left\vert \xi e^{-i2n\tau
}\right\rangle _{a},  \label{8}
\end{equation}%
here $\left\vert \alpha \right\rangle $ and $\left\vert \xi \right\rangle $
are the coherent state and squeezed vacuum state, respectively.

By using the configuration of Fig.1, assuming that mode $a$ is initially in
a coherent state $\left\vert \alpha \right\rangle _{a}$, choosing $\tau =\pi
$, and detecting the output of the beam splitter BS2, Gerry \cite{Gerry}
obtained following superpositions of coherent states%
\begin{equation}
\left\vert \psi _{odd-coh}^{even-coh}\right\rangle _{a}\sim \left\vert
\alpha \right\rangle _{a}\pm \left\vert -\alpha \right\rangle _{a},
\label{9}
\end{equation}%
which are called the even and odd coherent states, respectively. If we
assume that mode $a$ is initially in a squeezed vacuum state $\left\vert \xi
\right\rangle _{a},$ and choose $\tau =\pi /2$, then we will obtain the
following superpositions of squeezed vacuum states%
\begin{equation}
\left\vert \psi _{\pm }\left( \xi \right) \right\rangle _{a}\sim \left\vert
\xi \right\rangle _{a}\pm \left\vert -\xi \right\rangle _{a}.  \label{10}
\end{equation}%
Since $\left\vert \psi _{+}\left( -\xi \right) \right\rangle =\left\vert
\psi _{+}\left( \xi \right) \right\rangle $, and $\left\vert \psi _{-}\left(
-\xi \right) \right\rangle =-\left\vert \psi _{-}\left( \xi \right)
\right\rangle $, they can be named as the even-parity squeezed vacuum states
and the odd-parity squeezed vacuum states, respectively. The superpositions
of squeezed vacuum states have some novel features. As an example, we
discuss their photon number distributions. The probability amplitude for
having $n$ photons in these states is
\begin{equation}
\left\langle n|\psi _{\pm }\right\rangle \sim \left\langle n|\xi
\right\rangle \pm \left\langle n|-\xi \right\rangle .  \label{11}
\end{equation}%
It is well known that only even photons can exist in a squeezed vacuum
state, i.e.%
\begin{equation}
\left\langle 2m|\xi \right\rangle =\frac{1}{\sqrt{\cosh r}}\frac{\sqrt{%
\left( 2m\right) !}}{2^{m}m!}\left( -e^{i\varphi }\tanh r\right) ^{m},
\label{12}
\end{equation}%
\begin{equation}
\left\langle \left( 2m+1\right) |\xi \right\rangle =0,  \label{13}
\end{equation}%
where $\xi =re^{i\varphi }$, and $m=0,1,2,\cdots \cdots $. Therefore,
\begin{equation}
\left\langle 2m|\psi _{\pm }\right\rangle \sim \frac{1}{\sqrt{\cosh r}}\frac{%
\sqrt{\left( 2m\right) !}}{2^{m}m!}\left( e^{i\varphi }\tanh r\right) ^{m}%
\left[ \left( -1\right) ^{m}\pm 1\right] ,  \label{14}
\end{equation}%
it tells us that the states $\left\vert \psi _{+}\right\rangle $ can contain
only $n=2m=2(2k)=4k$ $\left( k=0,1,2,\cdots \right) $, i.e., $0,4,8,\cdots ,$
photons, while the states $\left\vert \psi _{-}\right\rangle $ can contain
only $n=2m=2(2k+1)=4k+2$, i.e., $2,6,10,\cdots ,$ photons.\bigskip\

Now let us describe the procedures of getting Eq.(10) in some details
(Fig.1). Assume the input state to BS1 is%
\begin{equation}
\left\vert \psi _{0}\right\rangle =\left\vert 1\right\rangle _{b}\left\vert
0\right\rangle _{c}\equiv \left\vert 10\right\rangle _{bc},  \label{15}
\end{equation}%
then the output state of BS1 reads
\begin{equation}
\left\vert \psi _{1}\right\rangle =\frac{1}{\sqrt{2}}\left( \left\vert
10\right\rangle _{bc}+i\left\vert 01\right\rangle _{bc}\right) ,  \label{16}
\end{equation}%
where Eq.(3) has been used. Assume that mode $a$ is initially in a squeezed
vacuum $\left\vert \xi \right\rangle _{a},$then the state before the Kerr
medium KM and the phase shifter $\theta $ is%
\begin{equation}
\left\vert \psi _{2}\right\rangle =\left\vert \xi \right\rangle
_{a}\left\vert \psi _{1}\right\rangle =\left\vert \xi \right\rangle _{a}%
\frac{1}{\sqrt{2}}\left( \left\vert 10\right\rangle _{bc}+i\left\vert
01\right\rangle _{bc}\right) ,  \label{17}
\end{equation}%
and the state after the Kerr medium and the phase shifter reads%
\begin{equation}
\left\vert \psi _{3}\right\rangle =\frac{1}{\sqrt{2}}\left( \left\vert \xi
e^{-i2\tau }\right\rangle _{a}\left\vert 10\right\rangle _{bc}+ie^{i\theta
}\left\vert \xi \right\rangle _{a}\left\vert 01\right\rangle _{bc}\right) ,
\label{18}
\end{equation}%
where Eq.(4) and Eq.(8) have been used. The output state of BS2 is
\begin{eqnarray}
\left\vert \psi _{4}\right\rangle  &=&\frac{1}{2}\left[ \left\vert \xi
e^{-i2\tau }\right\rangle _{a}\left( \left\vert 10\right\rangle
_{bc}+i\left\vert 01\right\rangle _{bc}\right) +ie^{i\theta }\left\vert \xi
\right\rangle _{a}\left( \left\vert 01\right\rangle _{bc}+i\left\vert
10\right\rangle _{bc}\right) \right]   \notag \\
&=&\frac{1}{2}\left( \left\vert \xi e^{-i2\tau }\right\rangle
_{a}-e^{i\theta }\left\vert \xi \right\rangle _{a}\right) \left\vert
10\right\rangle _{bc}  \notag \\
&&+\frac{1}{2}i\left( \left\vert \xi e^{-i2\tau }\right\rangle
_{a}+e^{i\theta }\left\vert \xi \right\rangle _{a}\right) \left\vert
01\right\rangle _{bc},  \label{19}
\end{eqnarray}%
where Eq.(3) has been used again. Choosing $\tau =\pi /2,$ and $\theta =0,$
we obtain%
\begin{equation}
\left\vert \psi _{4}\left( \tau =\frac{\pi }{2},\theta =0\right)
\right\rangle =\frac{1}{2}\left[ \left( \left\vert -\xi \right\rangle
_{a}-\left\vert \xi \right\rangle _{a}\right) \left\vert 10\right\rangle
_{bc}+i\left( \left\vert -\xi \right\rangle _{a}+\left\vert \xi
\right\rangle _{a}\right) \left\vert 01\right\rangle _{bc}\right] .
\label{20}
\end{equation}%
Now if detector $D_{b}$ fires, then mode $a$ will be projected into the
odd-parity squeezed vacuum states, and if detector $D_{c}$ fires, mode $a$
will be projected into the even-parity squeezed vacuum states of Eq.(10). By
choosing different values of $\tau $ and $\theta $, one can get different
kinds of superpositions of squeezed vacuum states.

It should be noted that the condition $\tau =K\left( l/v\right) =\pi $ is
required for getting Eq.(9), the superpositions of coherent states; while $%
\tau =\pi /2$ is required for getting Eq.(10), the superpositions of
squeezed vacuum states. Since, in general, the Kerr nonlinearity $K$ is
small, and therefore, a long nonlinear medium (large $l$) is needed for
getting a large phase shift $\tau $ \cite{Sanders}$.$But for a long
nonlinear medium, the dissipative effects will be great importance and will
cause a decoherence of the desired superposition states. In this aspect, the
generation of the superpositions of squeezed vacuum states are easier than
that of the superpositions of coherent states.

In the following we discuss the preparation of the entanglement of squeezed
vacuum states. For this reason we modify the configuration of Fig.1 to that
of Fig.2. In this case, the state after the Kerr medium KM1 and the phase
shifter is described by Eq.(18). Now let mode $b$ interacts with mode $%
a^{\prime }$ (in a squeezed vacuum state $\left\vert \eta =r^{\prime
}e^{i\varphi ^{\prime }}\right\rangle $ ) via the Kerr medium KM2. The
interaction Hamiltonian has the form of Eq.(5), but with $K$ and $\hat{n}%
_{a} $ replaced by $K^{\prime }$ and $\hat{n}_{a^{\prime }}$, respectively.
The states before and after the Kerr medium KM2 are%
\begin{equation}
\left\vert \psi _{5}\right\rangle =\frac{1}{\sqrt{2}}\left( \left\vert \xi
e^{-i2\tau }\right\rangle _{a}\left\vert 10\right\rangle _{bc}+ie^{i\theta
}\left\vert \xi \right\rangle _{a}\left\vert 01\right\rangle _{bc}\right)
\left\vert \eta \right\rangle _{a^{\prime }},  \label{21}
\end{equation}%
and%
\begin{equation}
\left\vert \psi _{6}\right\rangle =\frac{1}{\sqrt{2}}\left( \left\vert \xi
e^{-i2\tau }\right\rangle _{a}\left\vert \eta e^{-i2\tau ^{\prime
}}\right\rangle _{a^{\prime }}\left\vert 10\right\rangle _{bc}+ie^{i\theta
}\left\vert \xi \right\rangle _{a}\left\vert \eta \right\rangle _{a^{\prime
}}\left\vert 01\right\rangle _{bc}\right) ,  \label{22}
\end{equation}%
respectively. Here $\tau ^{\prime }=K^{\prime }t^{\prime }=K^{\prime }\left(
l^{\prime }/v^{\prime }\right) $. The output state from the BS2 then is%
\begin{eqnarray}
\left\vert \psi _{7}\right\rangle &=&\frac{1}{2}\left[ \left\vert \xi
e^{-i2\tau }\right\rangle _{a}\left\vert \eta e^{-i2\tau ^{\prime
}}\right\rangle _{a^{\prime }}\left( \left\vert 10\right\rangle
_{bc}+i\left\vert 01\right\rangle _{bc}\right) +ie^{i\theta }\left\vert \xi
\right\rangle _{a}\left\vert \eta \right\rangle _{a^{\prime }}\left(
\left\vert 01\right\rangle _{bc}+i\left\vert 10\right\rangle _{bc}\right) %
\right]  \notag \\
&=&\frac{1}{2}\left( \left\vert \xi e^{-i2\tau }\right\rangle _{a}\left\vert
\eta e^{-i2\tau ^{\prime }}\right\rangle _{a^{\prime }}-e^{i\theta
}\left\vert \xi \right\rangle _{a}\left\vert \eta \right\rangle _{a^{\prime
}}\right) \left\vert 10\right\rangle _{bc}  \notag \\
&&+\frac{1}{2}i\left( \left\vert \xi e^{-i2\tau }\right\rangle
_{a}\left\vert \eta e^{-i2\tau ^{\prime }}\right\rangle _{a^{\prime
}}+e^{i\theta }\left\vert \xi \right\rangle _{a}\left\vert \eta
\right\rangle _{a^{\prime }}\right) \left\vert 01\right\rangle _{bc}.
\label{23}
\end{eqnarray}%
Now if one of the detectors, $D_{c}$ or $D_{b},$ fires, then mode $a$ and
mode $a^{\prime }$ will be projected into the entanglement states%
\begin{equation}
\left( \left\vert \xi e^{-i2\tau }\right\rangle _{a}\left\vert \eta
e^{-i2\tau ^{\prime }}\right\rangle _{a^{\prime }}\pm e^{i\theta }\left\vert
\xi \right\rangle _{a}\left\vert \eta \right\rangle _{a^{\prime }}\right) ,
\label{24}
\end{equation}%
respectively. Different entanglement states can be obtained by choosing
different parameters $\tau $, $\tau ^{\prime }$, or $\theta $. It should
also be pointed out that the entanglement states between two coherent states
and that between one coherent state and one squeezed vacuum state can be
generated in a similar way, by choosing the input states of mode $a$ and
mode $a^{\prime }$ suitably.

Now let us consider experimental aspects. There are two most important
ingredients in the proposal: The Kerr nonlinearity and the single-photon
sources (SPS). As pointed out above, the Kerr nonlinearity required for
generating the superposition of squeezed vacuum states is only $1/2$ of that
required for generating the superposition of coherent states, this makes the
realization of the former is easier than the later. In addition, there are
some approaches to enhance the Kerr nonlinearity, for example, one can
enhance the Kerr nonlinearity by enclosing the medium in a cavity\cite%
{Agarwal}, or one can obtain giant Kerr nonlinearity from
electromagnetically induced transparency\cite{Schmidt}. On the other hand,
there are great progresses in the development of SPS in recent years, for
example, several deterministic SPSs have been realized in cavity quantum
electrodynamics and ion trap experiments\cite{Walther}\cite{Kimble}\cite%
{Rempe}.

In conclusion, we have proposed a scheme for generating the superposition
and entanglement of squeezed vacuum states. There are two most important
ingredients in the proposal, i.e. the cross Kerr nonlinearity and the single
photon sources (SPS). The studies on both of them have made great progress
in recent years, and this makes it possible to realize the proposal of this
paper in experiments. It should also be pointed out that the present
proposal can be extended to generate the entanglement states between two
coherent states and that between one coherent state and one squeezed vacuum
state.

This work was supported by the National Natural Science Foundation of China
under grant numbers 60578055 and 60178001.

Figure captions

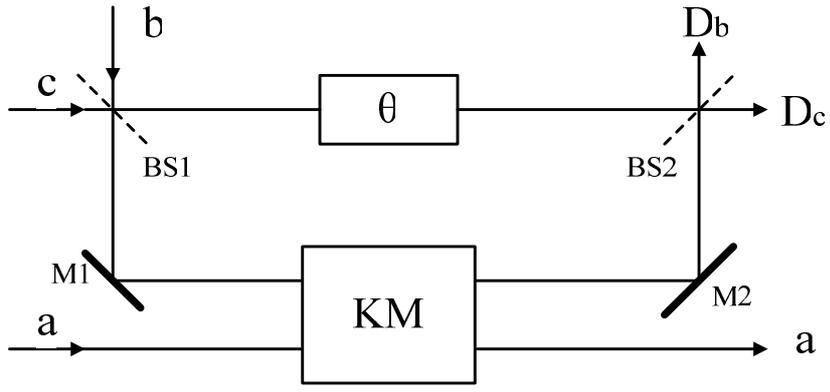
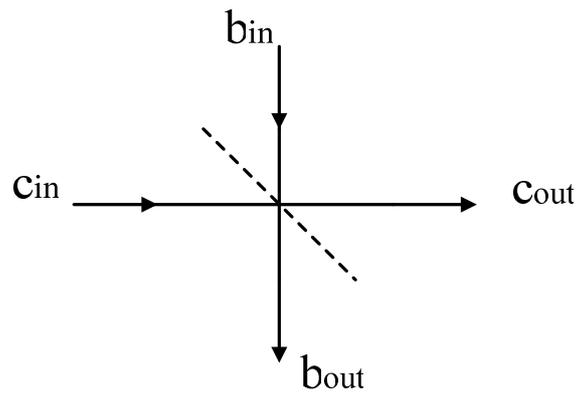
Fig.1. The configuration for generating the superpositions of squeezed
vacuum states. BS1 and BS2: beam splitters; KM: Kerr medium; $\theta $:
phase shifter; D$_{b}$ and D$_{c}$: photon detectors; M1 and M2: mirrors.

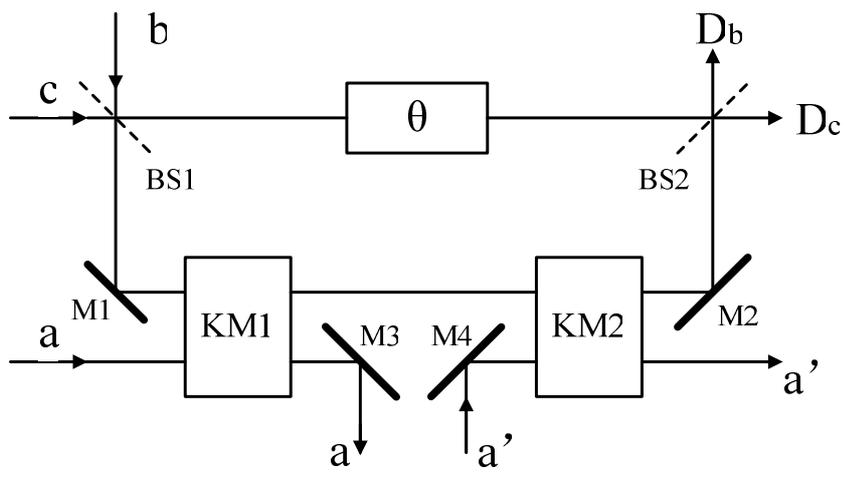
Fig.2. The configuration for generating the entanglements of squeezed vacuum
states. BS1 and BS2: beam splitters; KM1 and KM2: Kerr medium; $\theta $:
phase shifter; D$_{b}$ and D$_{c}$: photon detectors; M1-M4: mirrors.

Fig.3. The 50:50 beam splitter.

\end{document}